\documentclass[3p,times,procedia]{elsarticle}
\usepackage{nupha_ecrc}

\volume{00}

\firstpage{1}

\journalname{Nuclear Physics A}

\runauth{}

\jid{nupha}

\jnltitlelogo{Nuclear Physics A}

\usepackage{amssymb}

\usepackage[figuresright]{rotating}

\begin{document}

\begin{frontmatter}



\dochead{XXVIIth International Conference on Ultrarelativistic Nucleus-Nucleus Collisions\\ (Quark Matter 2018)}

\title{Heavy-flavour hadron decay leptons in Pb--Pb and Xe--Xe collisions at the LHC with ALICE}


\author{Andrea Dubla for the ALICE Collaboration}

\address{Physikalisches Institut,
  Universit{\"a}t Heidelberg, 69120 Heidelberg,
  Germany  \\ and GSI Helmholtzzentrum f{\"u}r
  Schwerionenforschung, 64248 Darmstadt, Germany}

\begin{abstract}
Heavy quarks, i.e. charm and beauty, are sensitive probes of the medium produced in high-energy heavy-ion collisions. They are produced in the early stage of the collision, mainly in hard partonic scattering processes, and are expected to experience the whole collision evolution interacting with the medium constituents via both elastic and inelastic processes.
The nuclear modification factor ($R_{\rm AA}$) is one of the main experimental observables that allow us to investigate the interaction strength of heavy quarks with the medium.
The ALICE collaboration measured the production of open heavy-flavour hadrons via their semi-leptonic decays to electrons at mid-rapidity and to muons at forward rapidity in elementary  proton-proton (pp) collisions as well as p--Pb, Pb--Pb  and in Xe--Xe collisions. 
Recent results will be discussed, and model calculations including the interaction of heavy quarks with the hot and dense QCD medium will be compared with the data.
\end{abstract}


\end{frontmatter}


\section{Introduction}
\label{}
The main goal of the ALICE experiment \cite{ALICE2} is to study the strongly-interacting matter at high energy density and temperature produced in ultra-relativistic heavy-ion collisions at the Large Hadron Collider (LHC).
In such collisions a deconfined state of quarks and gluons, the Quark-Gluon Plasma (QGP), is expected to be formed.
Due to their large masses, heavy quarks, i.e. charm ($c$) and beauty ($b$) quarks, are produced at the initial stage of the collision, almost exclusively in hard partonic scattering processes. Therefore, witnessing the full evolution of the system, they are effective probes to study the mechanisms of parton energy loss~\cite{Radiativea, Colla} and hadronisation~\cite{Recombinationa} in the hot and dense medium, giving insight on the QGP evolution and its transport coefficients.
Finally, initial-state effects due to the presence of a heavy nucleus in the collision
system can play a role. At low Bjorken-$x$ (below $10^{-2}$) the parton densities in nucleons 
bound in nuclei are reduced with respect to those in free nucleons. This so-called "shadowing" 
leads to a reduction of heavy-flavour production rates, becoming more pronounced with decreasing
$p_{\rm T}$~\cite{Eskola}. Initial-state effects and their impact on the nuclear modification factor are investigated in p--Pb collisions.
In order to exploit the sensitivity of heavy-flavour observables to medium effects 
a precise reference where such effects are not expected is needed and it is provided by pp collisions. In pp collisions, 
heavy-quark production can be described theoretically via perturbative QCD calculations
over the full quark momentum range, while such a description does not hold for gluon and light-quark production. Therefore, measurements of heavy-flavour production
cross sections in pp collisions are used to test perturbative QCD calculations and provide the necessary experimental reference for heavy-ion collisions.
The modification of the $p_{\rm T}$-differential yield in heavy-ion collisions with 
respect to pp collisions at the same centre-of-mass energy is quantified by the nuclear 
modification factor $R_{\rm AA}$, defined as:
 $R_{\rm AA}(p_{\rm T},y) = \frac{1}{\langle T_{\rm AA} \rangle} \cdot
\frac{{\rm d^{2}}N_{\rm AA}/{\rm d}p_{\rm T}{\rm d}y}{{\rm d^{2}}\sigma_{\rm pp}/{\rm d}p_{\rm T}{\rm d}y}$, 
where d$^{2}$$N_{\rm AA}$/d$p_{\rm T}{\rm d}y$ is the yield measured in heavy-ion collisions in a given $p_{\rm T}$ and $y$ interval, and 
d$^{2}$$\sigma_{\rm pp}$/d$p_{\rm T}{\rm d}y$ is the corresponding production cross section in pp collisions.

\section{Heavy-flavour hadron decay leptons in pp, Xe--Xe and Pb--Pb collisions}

\begin{figure}[htb]
\centering
\includegraphics[height=2.2in]{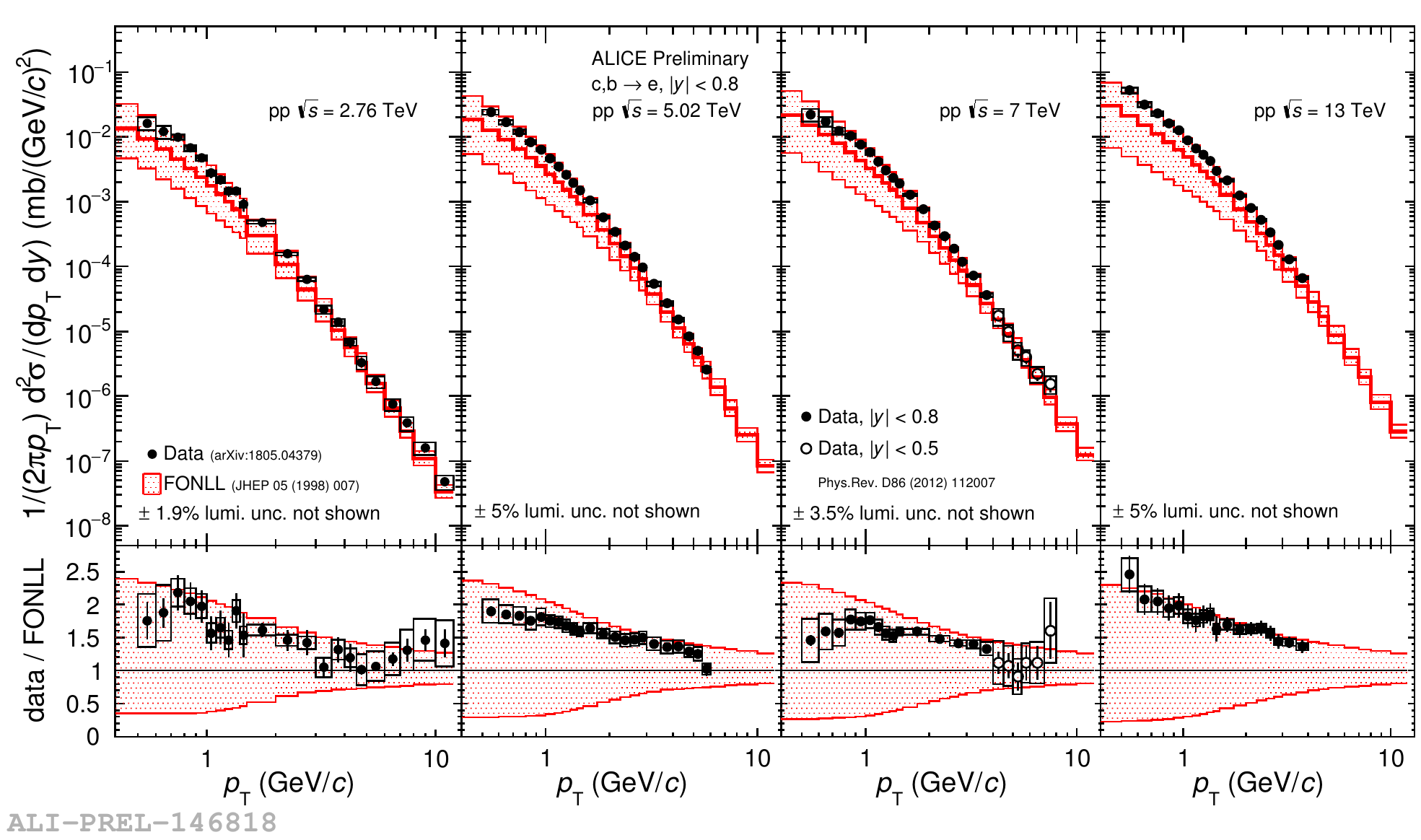}
\caption{$p_\mathrm{T}$-differential production cross section for electrons from heavy-flavour hadron decay in pp collisions at several collision energies (from left to right panel: $\sqrt{s}$ = 2.76, 5.02, 7 and 13 TeV) in comparison with FONLL calculations \cite{FONLL}. }
\label{fig1}
\end{figure}

The $p_{\rm T}$-differential production cross section of electrons from heavy-flavour hadron decays has been measured at mid-rapidity in pp collisions over a broad range of collision energies and it is shown in Fig. \ref{fig1} (from left to right panel: $\sqrt{s}$ = 2.76, 5.02, 7 and 13 TeV).
The photonic-tagging background subtraction method \cite{rpPbelectron} allowed for a reduction of the systematic uncertainties on the pp reference cross section by a factor of about 3 compared to the previously published cross sections at $\sqrt{s}$ = 2.76 and 7 TeV \cite{epp276, epp7}.
The dashed areas represent the FONLL calculation \cite{FONLL}. The full systematic band of the model originates from the variation of the factorization and normalization scales as well as the heavy-quark masses and the uncertainty on the parton distribution function (PDF) used. In the lower panels of Fig. \ref{fig1} the ratios of the experimental measurements with the central value of the FONLL calculations are shown. The measured cross sections are close to the upper edge of the FONLL uncertainty band for all the collision energies analysed.

\begin{figure}[htb]
\begin{minipage}[b]{0.4\linewidth}
\centering
\includegraphics[height=2.1in]{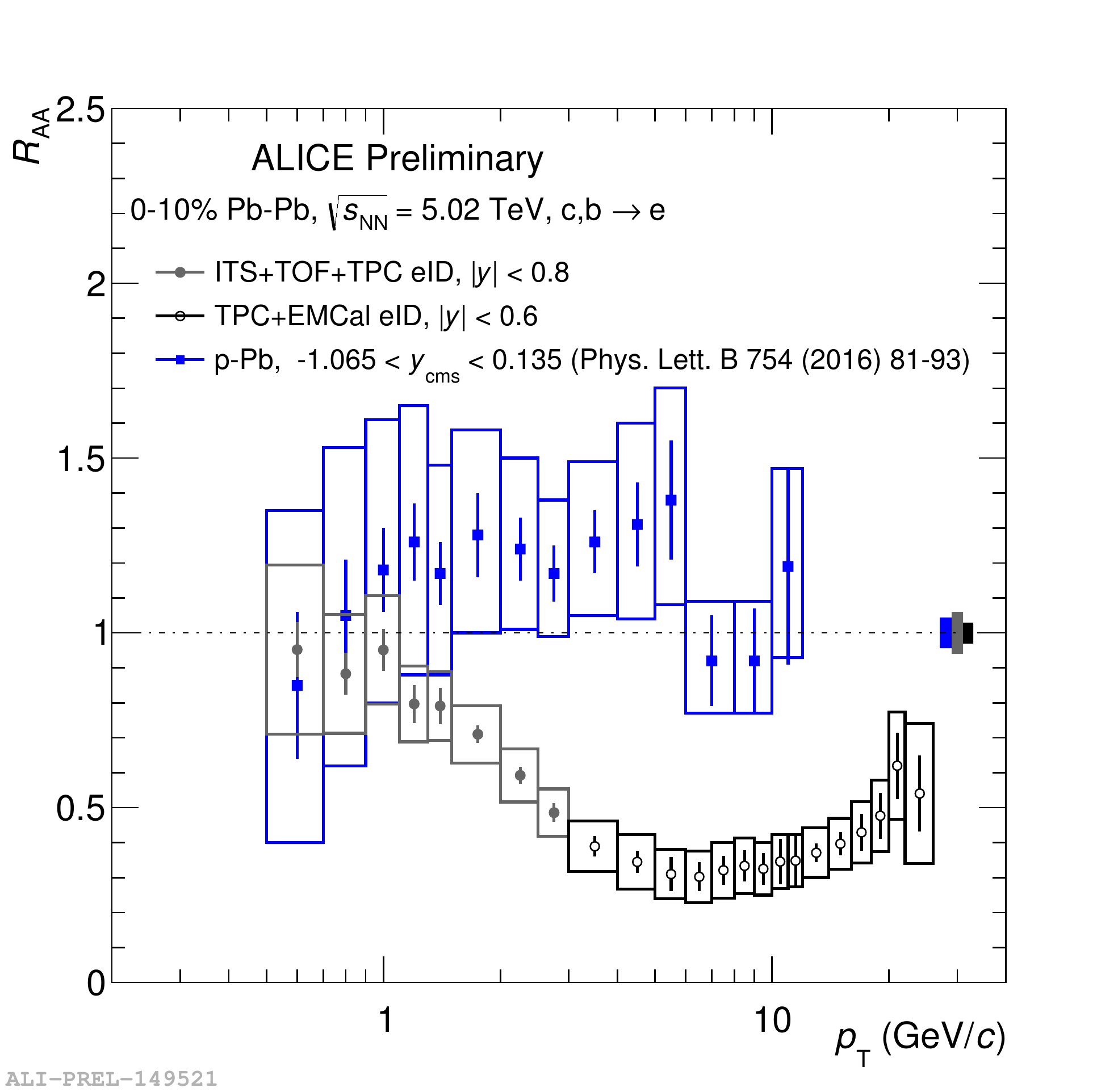}
\end{minipage}
\hspace{1cm}
\begin{minipage}[b]{0.5\linewidth}
\centering
\includegraphics[height=2.1in]{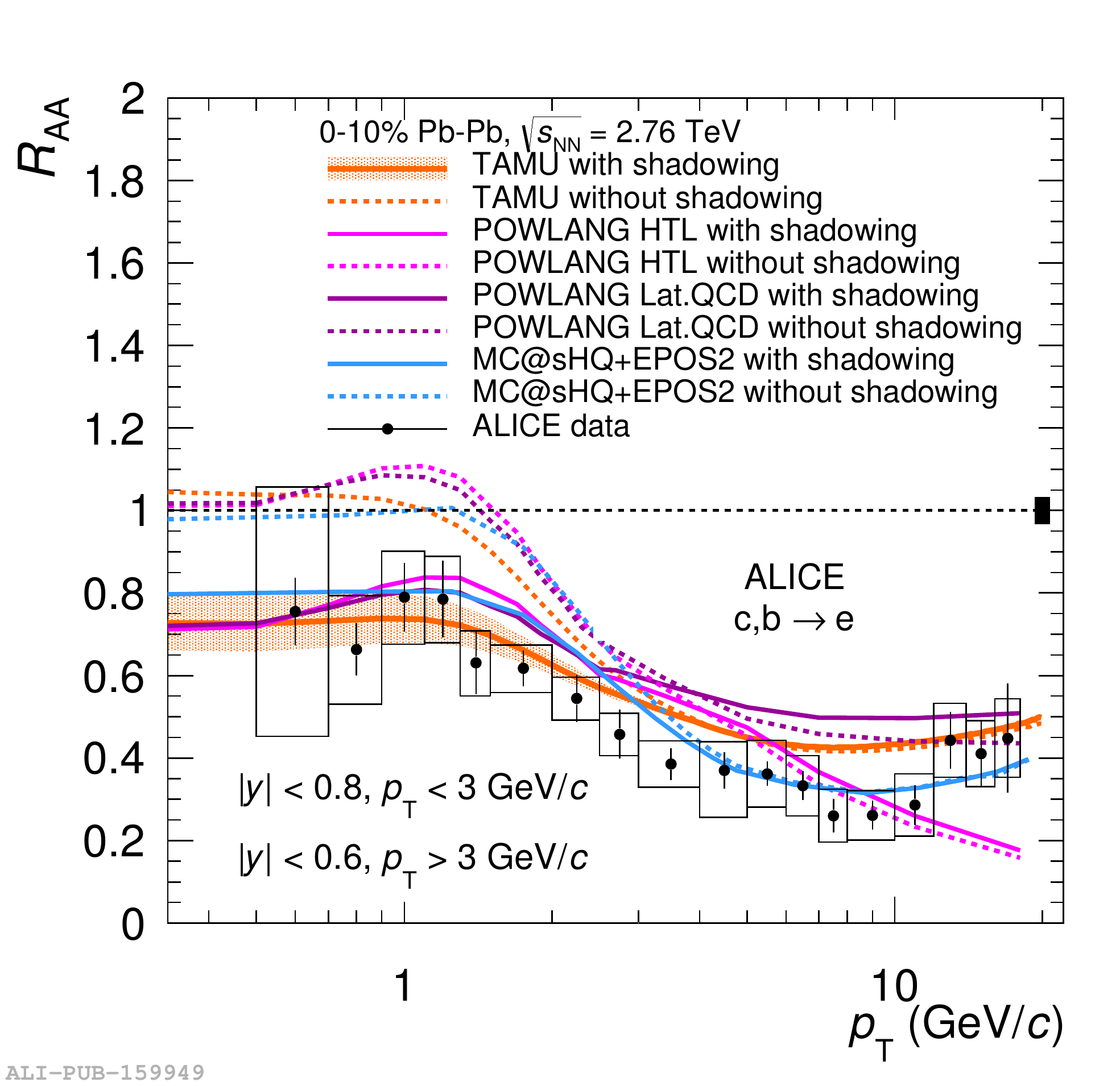}
\end{minipage}
\caption{ Left:  $R_{\rm AA}$ of electrons from heavy-flavour hadron decays as a function of $p_{\rm T}$ in central (0-10\%) Pb--Pb (black and grey markers) and in p--Pb (blue markers) collisions \cite{rpPbelectron} at $\sqrt{s}$ = 5.02 TeV. Right: $R_{\rm AA}$ of electrons from heavy-flavour hadron decays at  $\sqrt{s}$ = 2.76 TeV in comparison with model calculations with and without the inclusion of the EPS09 shadowing parameterisations \cite{POWLANG, EPOS, TAMU}.}
\label{fig2}
\end{figure}

The ALICE collaboration measured the $R_{\rm AA}$ of open heavy-flavour hadrons via their semi-leptonic decays in Pb--Pb collisions at $\sqrt{s_{\rm NN}}$ = 2.76 \cite{RAAelectron} and 5.02 TeV. The left panel of Fig. \ref{fig2}  shows the $R_{\rm AA}$ of electrons from heavy-flavour hadron decays as a function of $p_{\rm T}$ in central (0-10\%) Pb--Pb and in p--Pb collisions \cite{rpPbelectron} at $\sqrt{s}$ = 5.02 TeV. A strong suppression is observed in central Pb--Pb collisions above $p_{\rm T}$ = 1 GeV/$c$, while an $R_{\rm pPb}$ consistent with unity is measured, confirming that the suppression observed in central Pb--Pb collisions is predominantly induced by final-state effects due to the heavy quark energy loss in the medium.
In the low $p_{\rm T}$ region, the nuclear modification of the PDFs can play a significant role in the production of heavy quarks. This is addressed in the right panel of Fig. \ref{fig2}, which compares the measured  $R_{\rm AA}$ of electrons from heavy-flavour hadron decays at  $\sqrt{s}$ = 2.76 TeV with TAMU, POWLANG and MC@sHQ+EPOS2 model calculations  \cite{POWLANG, EPOS, TAMU} with and without the inclusion of the EPS09 shadowing parameterisations \cite{Eskola}. The depletion of the parton densities at low x, resulting in a reduced heavy-flavour production cross section per nucleon-nucleon pair in Pb--Pb collisions with respect to bare nucleon--nucleon collisions, leads to a reduction of the $R_{\rm AA}$ of electrons from heavy-flavour hadron decays at low $p_{\rm T}$. Data are better described when the nuclear PDFs are included in the theoretical calculations.

\begin{figure}[htb]
\begin{minipage}[b]{0.4\linewidth}
\centering
\includegraphics[height=2.in]{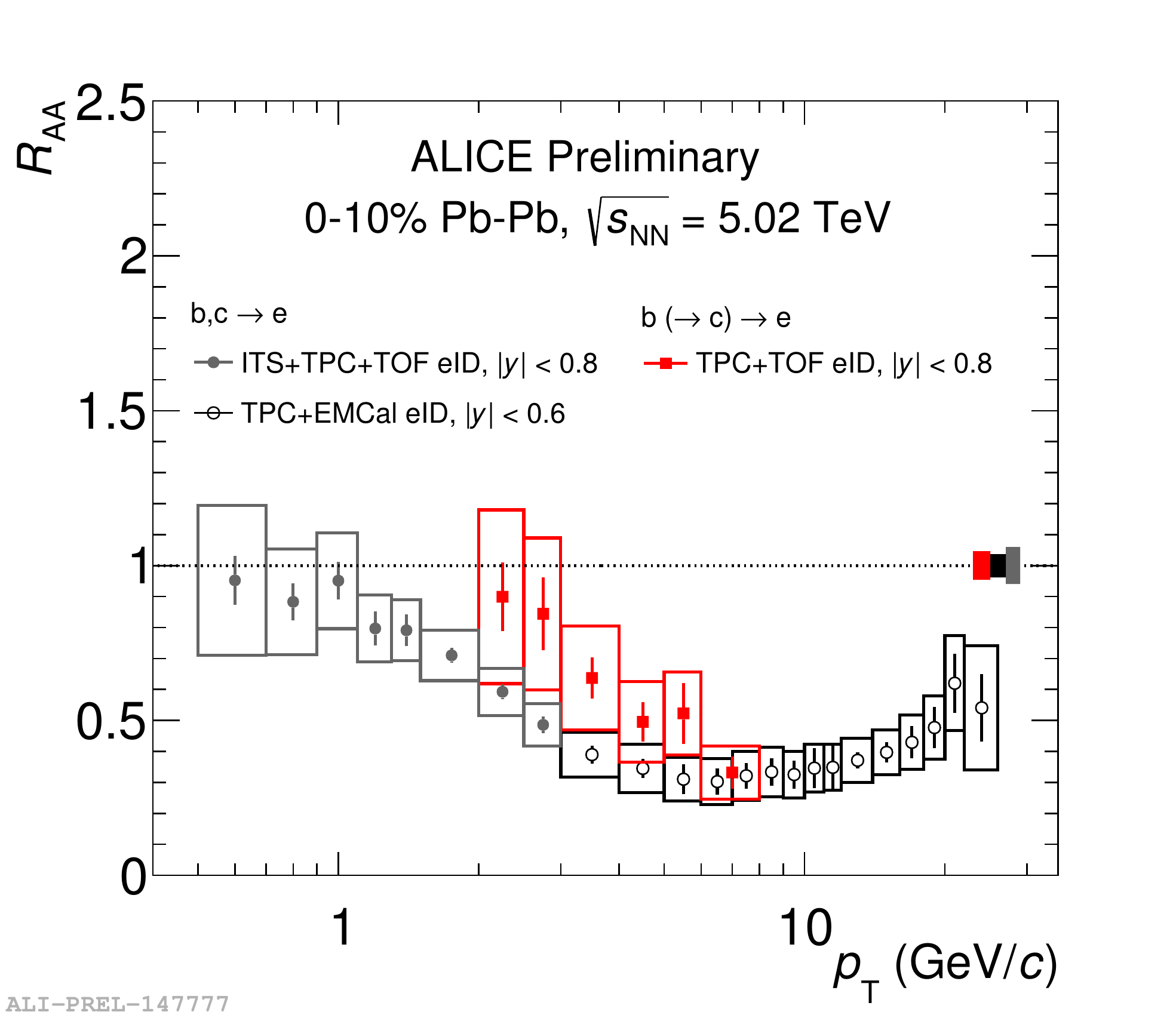}
\end{minipage}
\hspace{1cm}
\begin{minipage}[b]{0.5\linewidth}
\centering
\includegraphics[height=2.in]{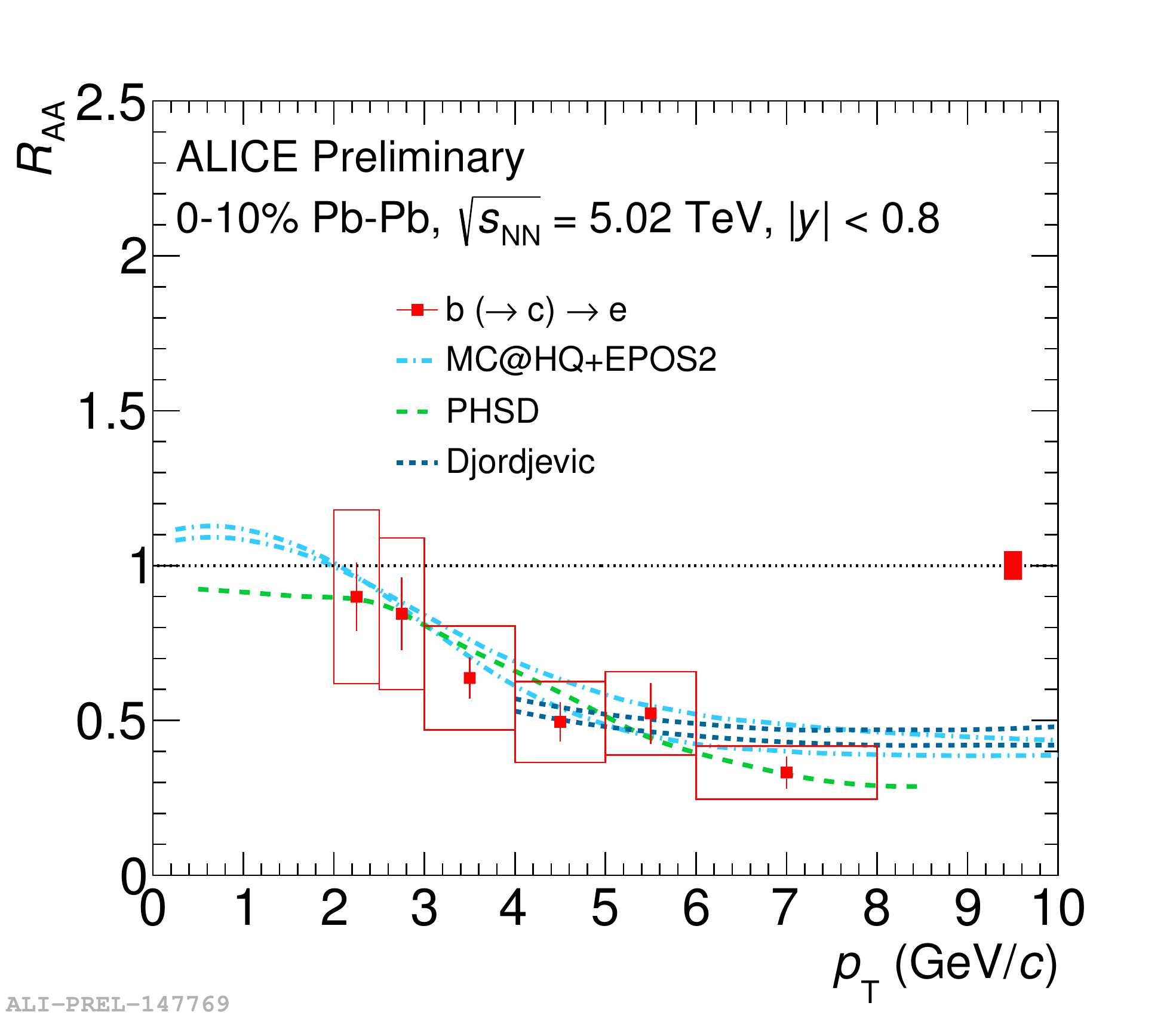}
\end{minipage}
\caption{ Left:  Comparison of the $R_{\rm AA}$
of electrons from beauty-hadron decays (red markers)  with  the  one  from  charm-  plus
beauty-hadron  decays (black markers) in Pb--Pb collisions at $\sqrt{s}$ = 5.02 TeV.   Right: $R_{\rm AA}$ of electrons from beauty hadron decay in comparison with models  implementing mass-dependent energy loss \cite{EPOS, Djordjevic, PHSD} in Pb--Pb collisions at $\sqrt{s}$ = 5.02 TeV.}
\label{figbeaury}
\end{figure}

Due to the QCD nature of parton energy loss, quarks
are predicted to lose less energy than gluons due to color-charge effect (smaller coupling with the plasma).  
In addition, the dead-cone effect \cite{DeadCone} is expected to reduce the energy loss
of massive quarks with respect to light quarks.  Therefore, if we consider only energy-loss effects, a hierarchy
in the $R_{\rm AA}$ is expected to be observed when comparing
the mostly gluon-originated light-flavour hadrons (e.g. pions) to D and to B mesons.
The measurement and comparison of these
different medium probes provide a unique test of the
colour-charge  and  mass  dependence of parton energy
loss.
A comparison of the $R_{\rm AA}$
of electrons from beauty-hadron  decays  with  the  one  from  charm-  plus
beauty-hadron  decays  is  shown  in the left panel of  Fig. \ref{figbeaury}  for  the
10\%  most  central  Pb--Pb  collisions at $\sqrt{s}$ = 5.02 TeV.   The  contribution
to the heavy-flavour decay electron yield due to beauty-
hadron  decays  was  extracted  by  means  of  a  fit  to  the
electron  impact  parameter  distribution. 
The two $R_{\rm AA}$ are fully compatible for $p_{\rm T}$ $>$ 6 GeV/$c$, where the beauty contribution is
larger than the charm one \cite{epp276}. At lower $p_{\rm T}$, where the charm contribution dominates, the data indicates a smaller suppression for electrons coming from beauty-hadron decays with respect to electrons from inclusive heavy-flavour hadron decays.
The measured $R_{\rm AA}$ is in agreement within uncertainties with models implementing mass-dependent energy loss \cite{EPOS, Djordjevic, PHSD}.

\begin{figure}[htb]
\centering
\includegraphics[height=2.2in]{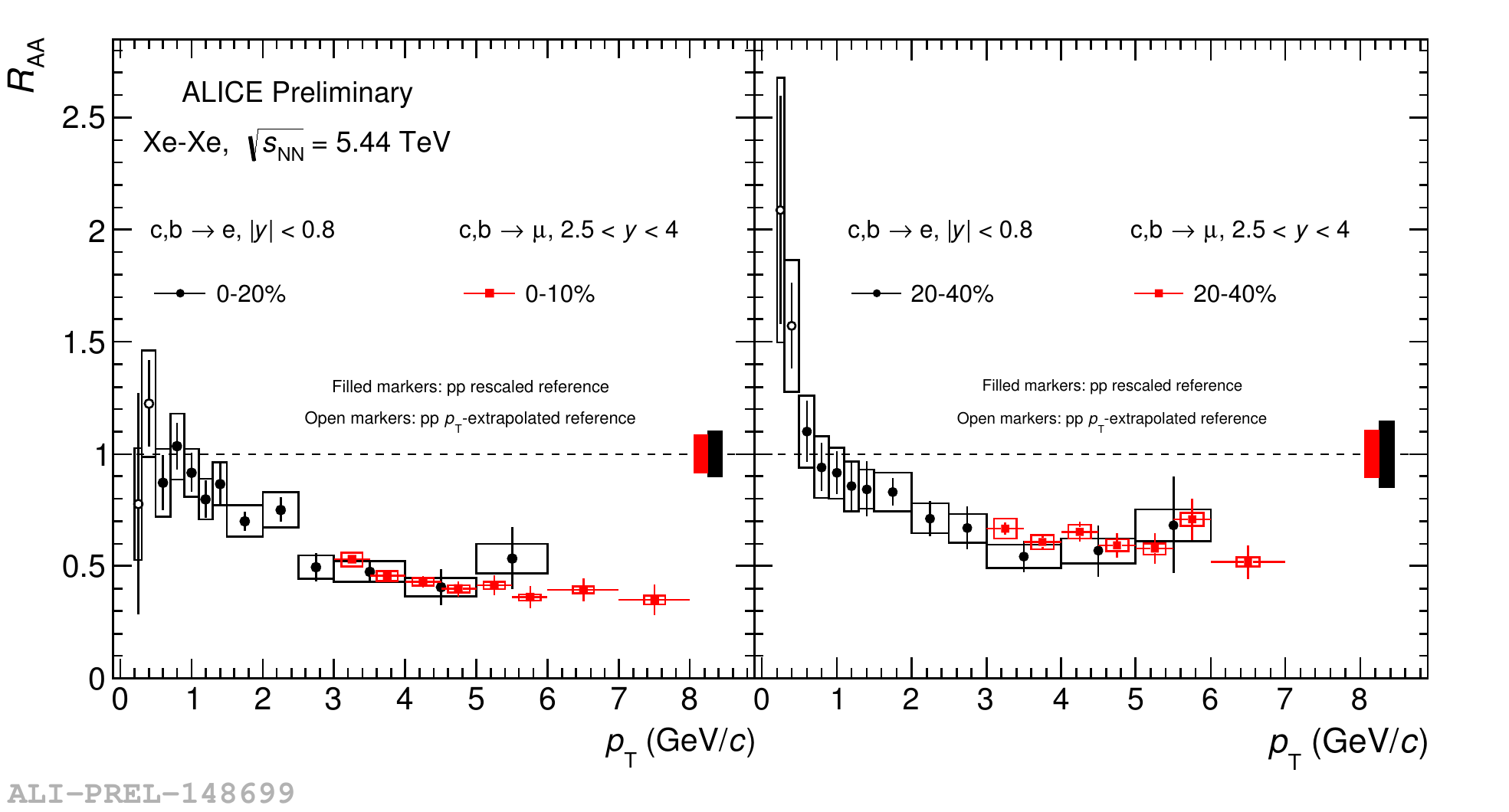}
\caption{Nuclear modification factors of electrons (black markers) and muons (red markers) from heavy-flavour hadron decay in central (left panel) and semi-central (right panel) Xe--Xe collisions at $\sqrt{s_{\rm NN}}$ = 5.44 TeV. }
\label{figxexe}
\end{figure}

The nuclear modification factor of electrons (and muons) from heavy-flavour hadron decays in Xe--Xe collisions at $\sqrt{s_{\rm NN}}$ = 5.44 TeV in the kinematic range 0.2(3) $<$ $p_{\rm T}$ $<$ 6(8) GeV/$c$ are reported for the centrality classes 0--20\% (0--10\%) and 20--40\% in the left  and right panel of Fig. \ref{figxexe}, respectively. A pp reference at $\sqrt{s}$ = 5.44 TeV is obtained by the extrapolation of the existing spectra at $\sqrt{s}$ = 5.02 TeV with a FONLL based scaling. The $R_{\rm AA}$ values of electron and muons from heavy-flavour hadron decays  are comparable in magnitude.
When comparing nuclear modification factors at similar ranges of averaged charged particle multiplicity densities, a remarkable similarity between central Xe--Xe collisions and Pb--Pb collisions at a similar center-of-mass energy is observed.
The comparison of the measured $R_{\rm AA}$ values in the two colliding systems could enable a test of the path length dependence of medium-induced parton energy loss.

\section{Conclusion}

Measurements of production of open heavy-flavour hadrons via their semi-leptonic decays to electrons at mid-rapidity and to muons at forward rapidity have been presented in pp, p--Pb, Pb--Pb and Xe--Xe collisions.
The results in Pb--Pb and in Xe--Xe collisions indicate a strong suppression of heavy-flavour decay leptons and of electrons from beauty-hadron decays with respect to elementary pp collisions. From the comparison  with  p--Pb  measurements  it  is  possible  to  conclude  that  the  suppression  observed  in  Pb--Pb and Xe--Xe collisions is mainly due to final state effects.


\bibliographystyle{elsarticle-num}
\bibliography{<your-bib-database>}



\end{document}